# Domain structures and superdislocations of $La_{0.7}Ca_{0.3}MnO_3$ thin films grown on $SrTiO_3$ substrates

Yong Ding and Jiaqing He[1]

Institut für Festkörperforschung, Forschungzentrum Jülich GmbH, D-52425 Jülich, Germany

**Abstract :** The domain structures and dislocations in epitaxial thin films of $La_{0.7}Ca_{0.3}MnO_3$ grown on $SrTiO_3$ substrates by pulsed laser deposition were investigated using Bragg-contrast diffraction and high-resolution transmission electron microscopy. It revealed that the films contained the $½[100]_o$ and $½[10\bar{1}]_o$ types partial threading dislocations, the 90°- and 120°- types of twin- domain boundaries, and two types $½[010]_o$, $½[111]_o$, of antiphase boundaries, which are often observed in bulk materials. In addition, two types of superdislocations were detected; one consisted of two $1/2[111]_o$ dislocations and a $1/2[111]_o$ antiphase boundary, and the other was composed of two $½[010]_o$ dislocations and a $½[010]_o$ antiphase boundary. These superdislocations, domain boundaries, and their relationships were extensively explored.

---

[1] Presently at The Center for Functional Nanomaterials, Bldg. 480, Brookhaven National Laboratory, Upton NY-11973, USA, email to jhe@bnl.gov.

## I. Introduction

The colossal magnetoresistance (CMR) manganite materials La$_{1-x}$Ca$_x$MnO$_3$, especially their thin films, have been extensively explored in recent years[1-17] because of their exciting potential practical applications and their interesting fundamental physics. In general, the properties of such thin epitaxial films strongly correlate with the defects within them, such as domain structures, precipitates, and threading dislocations, along with misfit dislocations at interfaces.[18] Misfits and threading dislocations have been widely studied in simple perovskite-oxide thin films, e.g., SrTiO$_3$, BaTiO$_3$, on different substrates.[18-21] In contrast, complex La$_{0.7}$Ca$_{0.3}$MnO$_3$ (LCMO) material is orthorhombic structure at room temperature. Thus, its basic vectors can be described as follows:

$$\begin{bmatrix} \mathbf{a}_o \\ \mathbf{b}_o \\ \mathbf{c}_o \end{bmatrix} \cong M \begin{bmatrix} \mathbf{a}_P \\ \mathbf{b}_P \\ \mathbf{c}_P \end{bmatrix} \quad (1)$$

Where M is a transformation matrix, equal to one of six matrixes $\begin{bmatrix} 0 & -1 & 1 \\ 2 & 0 & 0 \\ 0 & 1 & 1 \end{bmatrix}$, $\begin{bmatrix} 0 & 1 & 1 \\ 2 & 0 & 0 \\ 0 & 1 & -1 \end{bmatrix}$, $\begin{bmatrix} 1 & 0 & -1 \\ 0 & 2 & 0 \\ 1 & 0 & 1 \end{bmatrix}$, $\begin{bmatrix} 1 & 0 & 1 \\ 0 & 2 & 0 \\ -1 & 0 & 1 \end{bmatrix}$, $\begin{bmatrix} -1 & 1 & 0 \\ 0 & 0 & 2 \\ 1 & 1 & 0 \end{bmatrix}$, and $\begin{bmatrix} 1 & 1 & 0 \\ 0 & 0 & 2 \\ 1 & -1 & 0 \end{bmatrix}$; the subscripts *o* and *p* denote the indices for the orthorhombic cell, and for the perovskite primitive cell, respectively. Accordingly, the presence of the enlarged unit cell in the orthorhombic phase indicates the appearance of different types of orientation- and translation-domain structures[3,4]; For example, three types of antiphase boundaries (APBs), ½[010]$_o$, ½[101]$_o$, and ½[111]$_o$, along with 90°, 120° domain boundaries were identified in ceramics samples.[3,8] Also, some perfect dislocations seen in simple perovskite structures become partial ones in the LCMO orthorhombic structure. These partial dislocations usually form superdislocations in connection with the antiphase boundaries. Although there have many reports from domain-structure investigations in LCMO bulk and thin films[2-17], superdislocations are seldom mentioned, and to our knowledge, were only observed in ordered alloys.[22]

In this work, by means of using Bragg-diffraction contrast and high-resolution transmission electron microscopy, we investigated in detail the domains and dislocations of LCMO thin films deposited by pulsed laser on $SrTiO_3$ substrates. Our structural analyses shed light on superdislocations relative to the type and formation of dislocations and domains .

## II. Experiment

LCMO thin films, about 270nm, were deposited on a $SrTiO_3$ (STO) single crystal by pulsed laser deposition (PLD). The detailed procedure was reported elsewhere.[23] Specimens for transmission electron microscope (TEM) studies were prepared by a standard procedure. Plane-view specimens, parallel to the (100) STO plane, were made by thinning from the substrate side. They were first ground mechanically, then dimpled, and lastly, ion-beam-milled while cooling with liquid nitrogen. The diffraction patterns and Bragg-diffraction contrast images were recorded with a Philips CM20 electron microscope. High-resolution transmission electron microscopic (HRTEM) observations were carried out with a JEOL 4000EX microscope operated at 400 kV.

## III. Results

**A. Domain boundaries**

Figures 1(a) and (b) show two dark-field images of the plane-view LCMO sample obtained from the 212 and $01\bar{3}$ reflections, respectively. Figures 2 (a) and (b), representing the same areas as in Fig. 1, are another two dark-field pictures obtained from reflections $00\bar{2}$ and $20\bar{2}$, respectively. A comparison reveals that the lines in figure 1 have been extinguished in figure 2. Considering the extinction rules:

**g.R**=nπ (2)

where **g** and **R** are the reciprocal lattice vector and translation vector, respectively, and *n* is an integer [24]. The APBs will show contrast when n is an odd number in this formulation. Therefore, the curves denoted by the letters m and n in Fig. 1(a) should represent APBs that can be characterized by $1/2[111]_o$ or $1/2[010]_o$ translations. Combining the contrast image using the $01\bar{3}$ reflection (Fig. 1(b)) with the extinction rules, we judge that the boundary labeled as letter m is a 1/2[010] APB, and that marked by n is an 1/2[111] APB.

The domain boundaries between the dark and bright areas in Fig. 1(b) are identified as 90° domains boundaries since there are no relative reflections while the $01\bar{3}$ reflection rotates 90° around the $b_o$ axis. However, we only can confirm if one of the three types of 120° domains[3] exists in LCMO films. If it does, in Fig. 1(a) this kind of $120^0$ domain should show contrast. The other two types of 120° domains cannot be identified because, having a 120° rotation relationship, they will lose contrast due to the prohibition of 212 reflections rotated 120° around $[011]_o$. In Fig. 1(a), the main dark areas that we observed (marked as B and C) unfortunately represented precipitates of MnO. Their formation in the film's matrix is related to the presence a small excess of Mn reflecting a difference in the evaporation rate of different elements in the target during the film deposition (details are given in Reference 23). Accordingly, there is no evidence for the existence of the 120° domain boundaries in Fig. 1(a). However, the well-known key difference between the 120° domains is that their $b_o$ axes are perpendicular to each other.[3] To verify whether 120° domain boundaries actually exist, a selected-area diffraction pattern (SAED), shown in Figure 3, was obtained from the same area as Figure 1. This diffraction pattern can be classified as having $(010)_o$ and $(101)_o$ planes in reciprocal space due to the appearance of two superlattices, marked by arrows. The intensity of these superlattices indicates that, in the main part of the thin film, the $b_o$ axis is parallel to the normal plane of the film-substrate's interface. The superposition of $(010)_o$ and $(101)_o$ diffraction patterns reveals the existence of 120° domain boundaries. Since Fig 1 (a) could not separate the 120° domains due to the low contrast under those conditions, we obtained lattice images to highlight the detailed structure of the 120° domains. As shown in Figure 4, the dotted lines separate the image to two parts due to their having a different $b_o$ axis, as identified from the long translation period; one part (outside the dotted lines) representing the $b_o$ axis is parallel to the direction of the electron beam; the other part (inside the dotted lines), representing the $b_o$ axis, lies in a horizontal direction in the plane of film-substrate interface. Thus, the dotted lines between the two domains signify the 120° domain boundaries. We note that the inside area of this 120 domain is very small, about 20-50nm, and inevitably, some dislocations are imbedded in the boundaries.

**B. Dislocations**

**B1. Superdislocations**

Besides the precipitates in Figs. 1 and 2, we consider that the dark dots, labeled as A, typically represent threading dislocations. Using equation 2 (the extinction rules ) cannot tell us anything about their nature because the end-on dislocations show additional contrast arising from the surface relaxation of their strain field.[25] We identified the dislocations and their relationship with the APBs from lattice images. Comparing Fig. 2 (a) with Fig. 1(a), most dislocations appear to connect with the APBs, and some APBs terminate at dislocations. We termed two dislocations, together with an intervening APB, a superdislocation.[22] Judging from Fig,1 and Fig.2, the displacement vector $R_m$ of APB 'm' is 1/2 [010]. We designated $b_{m1}$ and $b_{m2}$ as the Burgers vectors of the two dislocations bonded with APB 'm' because the Burgers vector of a dislocation must be equal to the displacement vectors of the APBs attached to it, or to its modulus, a lattice vector.[22] Thus, $b_{m1}$, $b_{m2}$ = ± 1/2[010]$_o$. Since the $b_o$ axes of these domains are parallel to the normal plane of the film-substrate interface, as assessed from the diffraction pattern in Fig. 3, then the Burgers vectors of the dislocations must lie in the same direction as the dislocation lines. Similarly, from Fig.1 (a) and (b), we judge that the displacement vector of APB 'n' is ½[111]$_o$; thus the dislocation at its end has ½[111]$_o$ as its Burgers vector. Although it is hard to identify a superdislocation with ½[010]$_o$ dislocations and ½[010]$_o$ APBs along the [010]$_o$ direction from HRTEM images, we can confirm the existence of a superdislocation of the ½[111]$_o$ or ½[101]$_o$ type because the projection of ½[111]$_o$ dislocations into the (010)$_o$ plane is not zero, but ½[101]$_o$, Hence, we can identify the ½[111]$_o$ dislocations and ½[111]$_o$ APBs along [010]$_o$ from lattice images.

To investigate the apparent superdislocations in detail, we observed the same area by HRTEM. Figure 5 shows two typical lattice images containing superdislocations. In Fig. 5(a), two dislocations were detected with the same projected Burgers vectors, which can be classified as being ½[101]$_o$, if we accept $[100]_o$ and $[001]_o$ as the direction of the axis shown in the figure. Also, a ½[101]$_o$ projected displacement was observed across the boundary between the two dislocations. This superdislocation could be identified as a ½[101]$_o$ dislocation + ½[101]$_o$ APB + ½[101]$_o$ dislocation for the Burgers vector equal to the displacement of APB; alternatively, the ½[111]$_o$ dislocation + ½[111]$_o$ APB + ½[111]$_o$ dislocation for the ½[111]$_o$ projected to (010) plane is the same as $1/2[101]_o$. The projected Burgers vectors of the two dislocations in Fig. 5(b) are perpendicular to each other, and can be identified as $1/2[101]_o$ and $1/2[10\bar{1}]_o$ using the coordinates marked in the figure. The projected displacement of the APB between them is $1/2[101]_o$ or $1/2[10\bar{1}]_o$. The superdislocations in Fig. 5(b) may be a $1/2[101]_o$ dislocation + $1/2[101]_o$ APB + $1/2[10\bar{1}]_o$ dislocation, or a $1/2[111]_o$ dislocation + $1/2[111]_o$ APB + $1/2[11\bar{1}]_o$ dislocation, since their Burgers vectors are equal to the displacement of APB or the displacement of APB modulated by a lattice vector $[001]_o$. Distinguishing between them is difficult in these lattice images. However, the images in Fig. 1 indicate that the displacement of most of the APBs is $1/2[111]_o$, whilst 1/2[101] APBs are seldom observed. Accordingly, we concluded that the APBs in Fig. 5 terminate at 1/2[111] dislocations, whilst a few of them, such as the 'm' in fig. 1, terminate at 1/2[010] dislocations.

**B2. Partial dislocations**

In our plane-view samples, we also observed another two types of conventional dislocations, which do not connect with any type of APB, namely, the '**A**' dislocation pointed out in Figs. 1 and . 2. The total projected Burgers vector in the $(010)_o$ plane is identified as [100]$_o$, and ½[10$\bar{1}$]$_o$ by the Burgers circuit in the lattice images of Figure 6 (a) and (b), respectively. If the Burgers vector has a component in the $[010]_o$ direction, e.g., b= ½[210]$_o$, and ½[11$\bar{1}$]$_o$, then a $[010]_o$ or ½[111]$_o$ APB would be linked to this type of dislocation. However, no APBs terminate at it, thereby suggesting that its Burgers vectors must be [100]$_o$ and ½[10$\bar{1}$]$_o$. In Figure 6 (a), the dislocation likely dissociates into two ½[100]$_o$ partial dislocations that are slightly separated because the dissociation reaction is energetically favorable as the $|b|^2$ value decreases from 1 to 1/2.

## IV. Discussion

There are some previous HRTEM researches on domain boundaries in LCMO films on SrTiO$_3$ substrates[9-16] such as that of Aarts et al.[10] who observed APBs in a 6-nm thin film wherein the $b_o$ axis periodicity shifts over to $a_p$. Typically, the distance between two APBs is around 10nm. However, in their later work,[15,16] Zandbergen et al. pointed out that the existence of a twin boundary is another possible explanation for the APB contrast. In thicker films (30 nm),[10] they had observed domain boundaries, where the $b_o$ axes of domains are perpendicular to each other. These boundaries belong to 120° domain walls, as demonstrated by space-group analysis, for one domain can transform into another by rotating 120° around ½[012]$_o$. Lebedev et al,[9, 11,] and Van Tendeloo et al.,[12] systematically investigated 250nm thin films grown at different temperatures; they deduced that the column contrast in the thin film came from the 90° domains. The observed APBs with displacements of $1/2[100]_o$ were non-conservative ones with an additional MnO layer inserted between the $\langle 100 \rangle_p$ planes. In our case, the 90° domain wall has a high density and always is combined with $1/2[111]_o$ APBs. However, it is difficult to separate the $[100]_o$ and $[001]_o$ directions in the HRTEM images along the $b_o$ axis, and accordingly, problematic to identify 90° domain walls from $[010]_o$ lattice images. Using the dark-field technique, resolves this problem, as is evident from Fig.1. The domains have common $b_o$ axes for the largest misfit between the $1/2 b_o$ and $a_p$ of cubic SrTiO$_3$. Thus, $[100]_o$ and $[010]_o$ are a little bit larger and smaller, respectively, than the $a_p$ of SrTiO$_3$.[13-15] Therefore, the 90° domain boundaries perpendicular to the interface will have a role in releasing the strain.

Dislocations sometimes seriously affect the properties of thin films, including misfit dislocations near the interface, and threading dislocations in the film. As described, we observed threading-type dislocations in LCMO films. In epitaxial simple perovskite oxides, e.g., $BaTiO_3$, $(Ba_{1-x}Sr_x)TiO_3$, and $SrTiO_3$, they are classified into two types, $[100]_p$ and $[110]_p$.[18-21] Further, both types usually dissociate into two partials to lower the energy. However, the situation in LCMO thin films is more complex due to the tilting of the oxygen octahedra .[2] As Amelinckx mentioned,[22] if an ordered crystal contains dislocations, which are perfect for a disordered crystal, but not for an ordered structure, an APB terminating at the dislocation is necessarily generated. The displacive phase-transition of LCMO films has the same consequence as the order-disorder phase transition in alloys, namely, both lead to enlarged unit cells. So, in LCMO thin films, some threading dislocations, as in simple perovskite-oxide films, become superdislocations ( Figs. 1 and 5) However, the $[100]_o$ or $[001]_o$ type dislocations are still perfect (Fig. 6), and while they usually dissociate into two ½[100]$_o$ partials as in simple perovskite-oxide films, they do not dissociate while such dislocations are linked by APBs. Seemingly, the APB breaks up the large stress in the dislocation's core.

Comparing Figs. 1 and 2, we find that most of the dislocations are embedded within the domain boundaries. Hence, their formation can be considered as the product of island growth processes in the thin film. The mismatch between the inter-atomic spacing on the (100) surface of $SrTiO_3$ and on the $(010)_o$ plane of LCMO also can produce small-angle misorientations among LCMO islands. When the islands coalesce to form a continuous film, the stress accumulated in the film is partially released by the generation of dislocations at the boundaries between the islands. Whilst the LCMO film is cooling down to room temperature, the $[100]_o$ and $[001]_o$ directions of each island are adjusted, so to form 90° domain boundaries to release the thermal expansion. At the same time, the $[100]_o$ dislocations can dissociate into two $1/2[100]_o$ partials, as described in BaTiO$_3$ and (Ba$_{1-x}$Sr$_x$)TiO$_3$ systems,[18-21] whilst a $1/2[010]_o$ APB will be introduced between two adjacent $1/2[100]_o$ dislocations. In the presence of precipitates, some strains in the LCMO film will be released; sometimes, one end of the APBs can terminate at the interface between an LCMO grain and a precipitate.

## V. Conclusions

The domain structures and dislocation types in 270nm LCMO thin films grown on SrTiO$_3$ (100) substrates by PLD were analyzed by traditional TEM and HRTEM techniques. The main plane-defects identified were 90° domain walls and $1/2[111]_o$ APBs, although 120° domain boundaries and $1/2[010]_o$ APBs were also observed. Considerable numbers of the 90°domain boundaries are combined with $1/2[111]_o$ APBs. Two types of superdislocations were identified; one composed of two $1/2[010]_o$ dislocations and a $1/2[010]_o$ APB, and another made up of two $1/2[111]_o$ dislocations and a $1/2[111]_o$ APB. Besides such superdislocations, two types ½$[100]_o$ and ½$[10\bar{1}]_o$ partial dislocations were detected. Most of these dislocations, embedded in the domain boundaries, are believed to be the product of the growth processes of islands in the thin films.

## Acknowledgements

The authors thank K. Urban and C. L. Jia for their helpful discussions and J. Schubert for preparing the La$_{0.7}$Ca$_{0.3}$MnO$_3$ film samples, and A.D. Woodhead for critical reading of the manuscript.10

**Figure captions**

Figure 1. Two dark-field images formed using the reflections (a) 212, and (b) $01\bar{3}$. A marks a $[100]_o$ dislocation. The letters n and m label APBs with a displacement of $1/2[111]_o$ and $1/2[010]_o$, respectively.

Figure 2. Two dark-field images using the reflections (a) $00\bar{2}$, and (b) $20\bar{2}$. The areas denoted by C and D are precipitates. A is the same dislocation as shown in Fig. 1.

Figure 3. Selected-area diffraction pattern from the same area as that of Figs. 1 and 2. It can be indexed as the $(010)_o$ and $(101)_o$ planes in reciprocal space, corresponding, respectively, to the arrows marking the center and side superlattice reflections.

Figure 4. A lattice image shows a 120° domain boundaries, indicated by the dotted lines. The $[010]_o$ direction of the middle domain can be identified from its long translation period.

Figure 5. Lattice images along $[010]_o$ direction. (a) and (b) show that a superdislocation formed by $1/2[111]_o$ APBs can terminate at two $1/2[111]_o$ dislocations, and at an $1/2[111]_o$ and a $1/2[11\bar{1}]_o$ dislocation, respectively.

Figure 6. Lattice images along $[010]_o$. (a) a $[100]_o$ dislocation dissociates into two ½ $[100]_o$ partials, and, (b) a dislocation with a Burgers Vector $1/2[10\bar{1}]_o$.

**Figure 1**

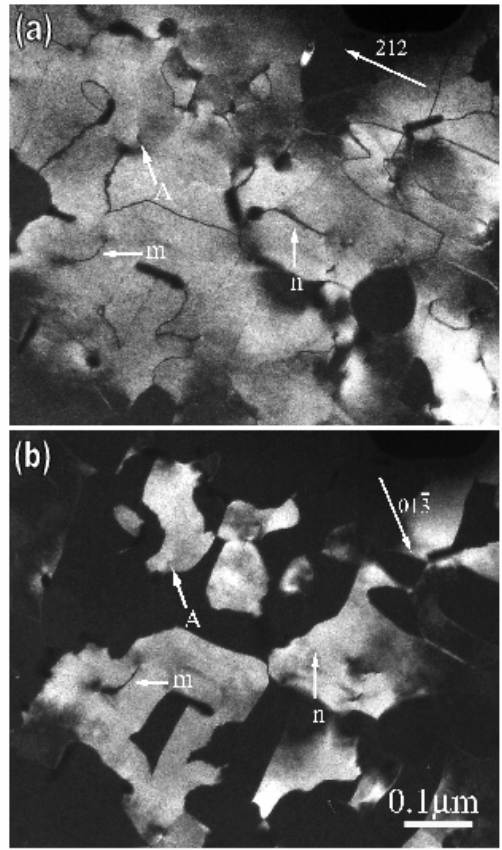

Figure 2

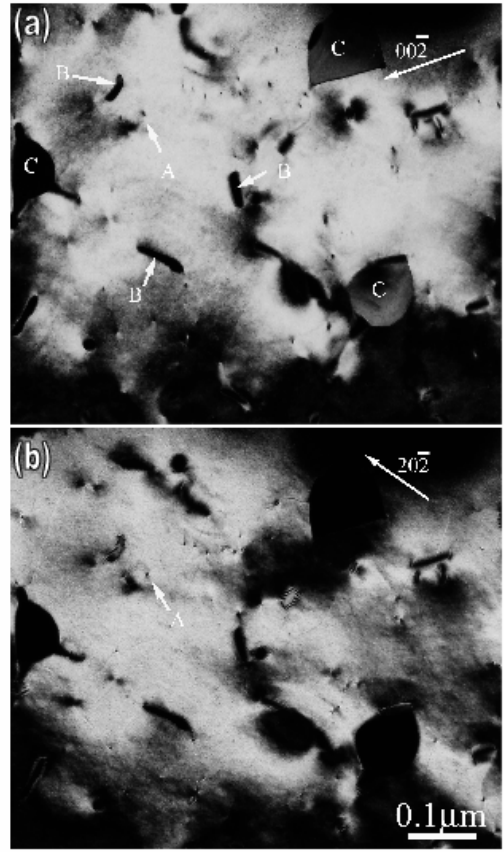

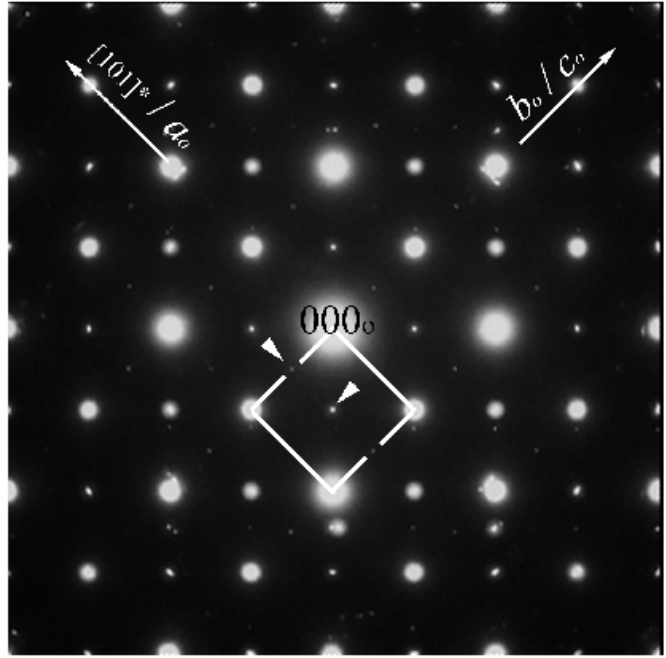

**Figure 4**

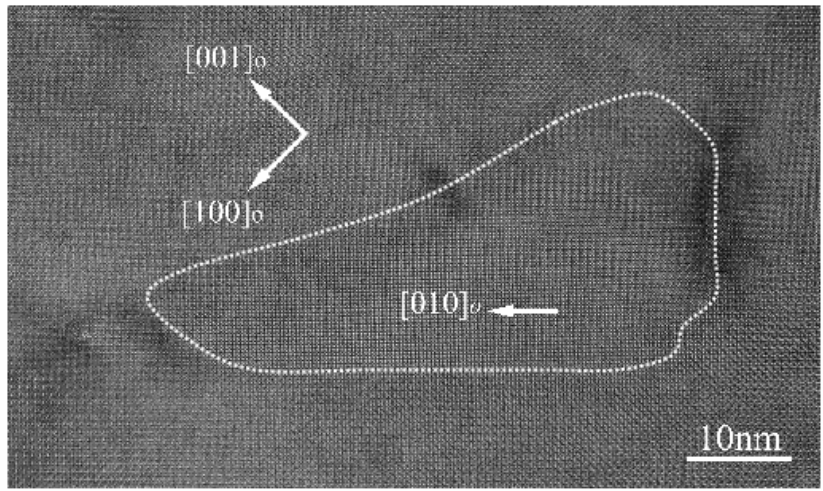

**Figure 5**

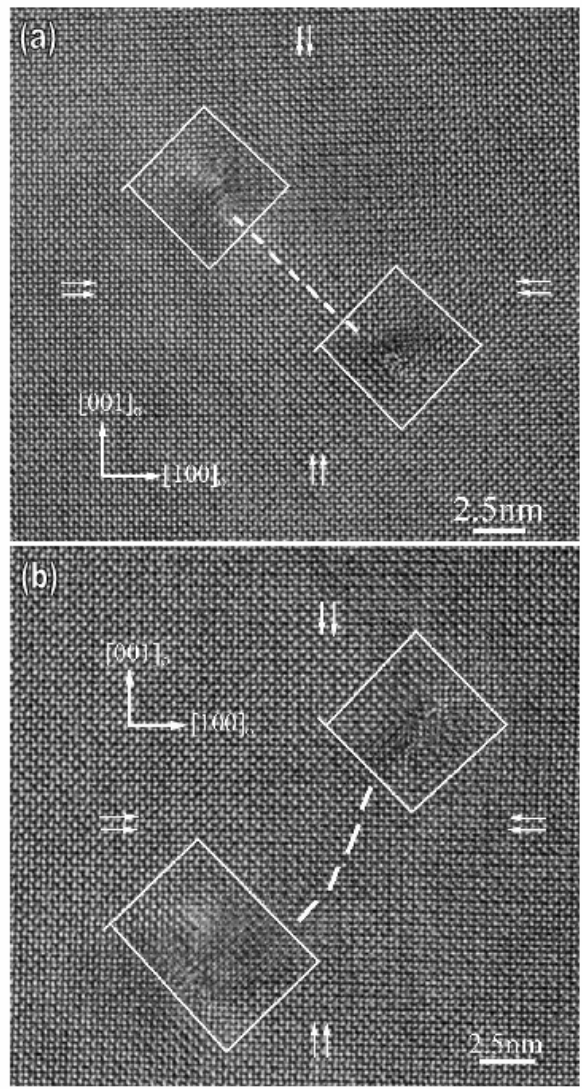

**Figure 6**

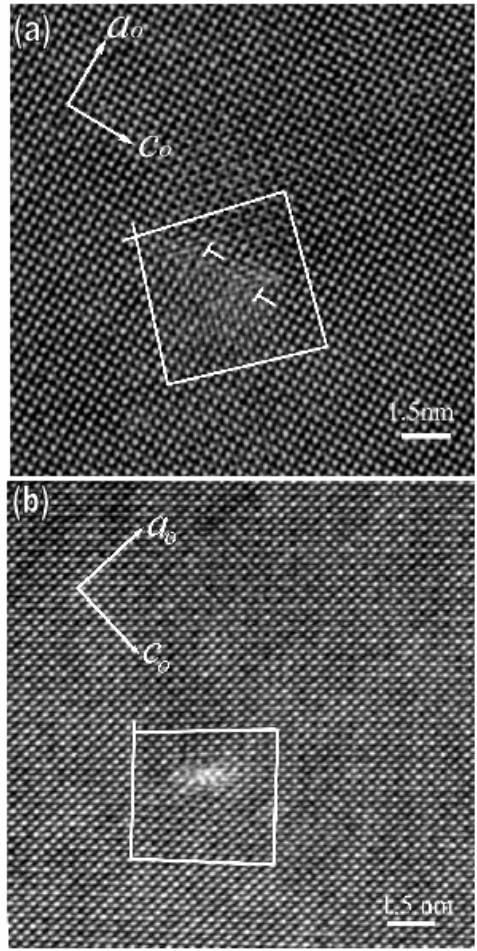